# Entropy Decrease in Isolated Systems with an Internal Wall undergoing Brownian Motion

### B. Crosignani[(*)]
*California Institute of Technology, Pasadena, California 91125*

and

### P. Di Porto
*Dipartimento di Fisica , Universita' dell'Aquila, 67010 L'Aquila, Italy and Istituto Nazionale di Fisica della Materia, Unita' di Roma 1, Roma, Italy*

*We consider an isolated gaseous system, divided in two parts by an adiabatic movable frictionless internal wall undergoing Brownian motion. We show how this kind of motion can lead to a substantial decrease of the system entropy. This surprising result does not violate Boltzmann's H-theorem since the molecular chaos assumption underlying H-theorem is not fulfilled by our system.*

It is generally assumed that -in accordance with the Second Law of thermodynamics- the entropy of an isolated system cannot decrease. The Second Law finds its most natural conceptual basis in Boltzmann's H-theorem, according to which complete *molecular chaos* is a sufficient condition for its validity [1]. In the present paper, we will consider a particular isolated system for which the standard molecular chaos assumption is not true and we will show how this system can undergo a relevant entropy decrease, thus violating the Second Law.

We refer to the case of an adiabatic cylinder containing 2n moles of the same perfect gas separated into two parts A and B (each containing n moles) by a movable, frictionless, *adiabatic* piston [2]. The piston is held by latches and the initial equilibrium states in A and B are respectively characterized by temperatures $T_1(0) + T_A$ , $T_2(0) + T_B$ , and volumes $V_A = SX(0)$, $V_B = S[L-X(0)]$ , where S and L are the transverse area and the length of the cylinder, and $X(0)$ labels the piston position. At time t=0, we release the latches and we assume the piston to slide in the cylinder without friction (see Fig.1).

**Fig.1** Adiabatic cylinder with adiabatic piston. (a) Piston held by latches. (b) Sliding piston at a generic time t.

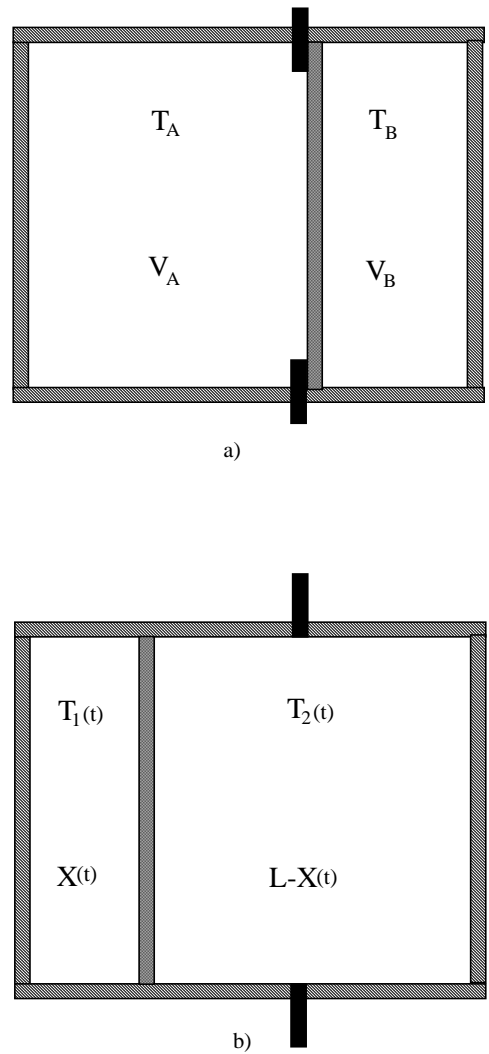

The problem of predicting the final equilibrium state of the system is indeterminate from a purely thermodynamic point of view. Actually, equilibrium thermodynamics only allows one to determine the common final pressure $P_f$ of the two gases. In fact, the equality of the final equilibrium pressures in A and B yields $P_f=nR(T_{1f}/V_{1f})=nR(T_{2f}/V_{2f})$, where $T_{1f}$, $V_{1f}$ and $T_{2f}$, $V_{2f}$ are the final temperatures and volumes in A and B, respectively, while energy conservation for a perfect gas is expressed by the relation $T_{1f}+T_{2f}=T_A+T_B$. From these two relations it is immediate to deduce $P_f=nR(T_{1f}/V_{1f})(1+T_{2f}/T_{1f})/(1+V_{2f}/V_{1f})$, so that the final pressure is given by $P_f=nR(T_A+T_B)/(V_A+V_B)$, but $T_{1f}$, $V_{1f}$, $T_{2f}$, $V_{2f}$ cannot be determined separately. In other words, the considered system is peculiar in that one cannot predict the final temperatures and volumes by means of elementary thermodynamics. This is related to the presence of the infinite set of contiguous equilibrium states obeying the above relations.

While the indetermination of final volumes and temperatures holds in the realm of elementary thermodynamics, a recent dynamical analysis [3] shows the possible existence of a well-defined final equilibrium point dependent on the initial conditions and allows one to investigate the dynamical evolution toward this point. More precisely, under the assumption that the two gases in A and B achieve thermal equilibrium over a time scale negligible with respect to that of the piston motion, the instantaneous state of the system is described by the set of evolution equations [3]

$$nc_v\dot{T}_1 \;=\; -\; nRT_1(\dot{X}/X) + \sqrt{8nRM_g/\pi}\sqrt{T_1}(\dot{X}^2/X) - M_g(\dot{X}^3/X) \;, \;(1)$$

$$nc_v\dot{T}_2 \;=\; nRT_2[\dot{X}/(L\text{-}X)] \;+\; \sqrt{8nRM_g/\pi}\sqrt{T_2}[\dot{X}^2/(L-X)]$$

$$+\; M_g[\dot{X}^3/(L\text{-}X)] \;, \qquad (2)$$

$$\ddot{X} \;=\; (nR/M)[(T_1/X) - T_2/(L-X)] - \sqrt{8nRM_g/\pi M^2}\;[(\sqrt{T_1}/X)$$

$$+\; \sqrt{T_2}/(L\text{-}X)]\dot{X} \;+\; (M_g/M)[\,(1/X) - (1/(L-X)]\dot{X}^2 \;, \;(3)$$

where $T_1$, $T_2$ and X label the temperature in A, the one in B and the piston position as functions of time, and the quantities R, $c_v$, $M_g$ and M represent the molar gas constant, the molar heat at constant volume, the common value of the gas masses in A and B, and the piston mass, and the dot denotes derivation with respect to time.

The set of Eqs.(1), (2), and (3) implies that piston position and temperatures reach asymptotic values $X_f$, $T_{1f}$, $T_{2f}$ dependent on the initial conditions, such that $T_{1f} + T_{2f} = T_A + T_B$ (energy conservation) and $T_{1f}/X_f = T_{2f}/(L-X_f)$ (equal final pressure in A and B).

We note that our dynamical model described by the coordinates X(t), $T_1(t)$, $T_2(t)$ does not lend itself to a Hamiltonian description. In order to get a physical insight about this statement, let us vary in all possible ways the initial temperature $T_A$ and, in correspondence, $T_B$, so to maintain the fixed value $T_A + T_B + 2T_0$. In this way, the initial conditions determined by X(0), $T_A$ span the square depicted in Fig.2, while the asymptotic values $X_f$, $T_{1f}$ lie on the diagonal , corresponding to the equilibrium pressure $P_f$ (see Eq.(3)). Therefore, an initial phase-space surface reduces to a final attractor of vanishing area, so that the phase-space area is not conserved as it should be for a Hamiltonian system (Liouville's theorem).

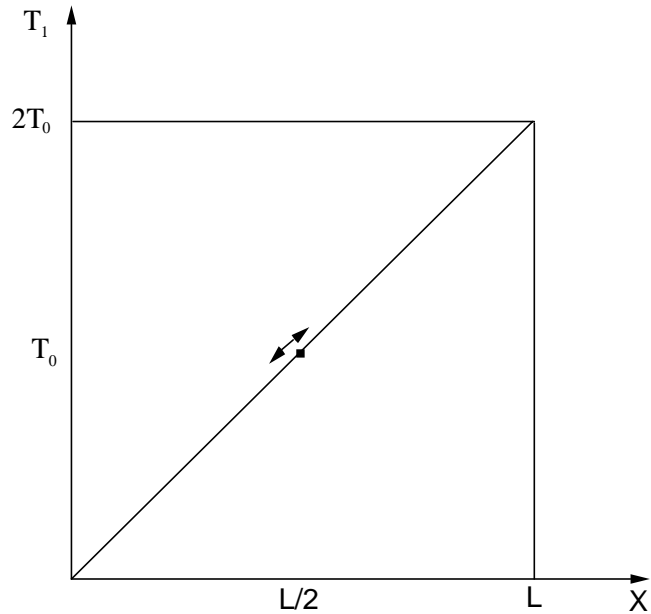

**Fig.2** Wandering of the state along the attractor (*quasi* - equilibrium line) starting from the maximum entropy state (L/2,$T_0$).

The above discussion refers to the dynamical evolution of the system toward a point on the *equal-pressure* equilibrium line (i.e., the attractor corresponding to the diagonal of Fig. 2), starting from an initial nonequilibrium state. In this paper, we wish to consider the further system evolution that takes place *starting from a point on the equilibrium line*. This evolution, associated with an extremely slow Brownian-like motion of the piston, essentially corresponds to a random wandering of the gas state in A ( and B) along the equilibrium line. This wandering is made possible by the fact that, due to pressure equality in A and B, the piston is acted upon by a very weak restoring force. In the following, we will place into evidence how, during this process, *the total entropy of the isolated system can substantially decrease*.

We will consider, in particular, the case in which the initial piston position is X(0)=L/2 and the initial common temperature in A and B is $T_0=(T_A+T_B)/2$ , corresponding to maximum entropy. During the transformation from this state to a final one corresponding to a generic X along the equilibrium line, both gases maintain the common constant pressure $P_f$, so that $T_1/X=T_2/(L-X)=2T_0/L$. The total entropy change $\Delta S(X)= \Delta S_1(X)+ \Delta S_2(X)=S(X)-S(L/2)$ is given by

$$\Delta S(X) = nc_p \left( \int_{T_0}^{T_1(X)} \frac{dT}{T} + \int_{T_0}^{T_2(X)} \frac{dT}{T} \right) = nc_p \left( ln\frac{T_1(X)}{T_0} + ln\frac{T_2(X)}{T_0} \right)$$

$$= nc_p \left( ln\frac{2X}{L} + ln\frac{2(L-X)}{L} \right) = nc_p \, ln\frac{4LX - 4X^2}{L^2} \leq 0 \quad , (4)$$

where $c_p$ is the molar heat at constant pressure. We wish now to obtain an estimate for the value attained by $<X^2>$ in an asymptotic time $t_{as}$. Once inserted into Eq.(4), we will show how it can give rise to a value of $|\Delta S|$ much larger than that associated with ordinary thermal entropy-fluctuations, that is the ones present in a gas at thermal equilibrium where any departure from equilibrium itself is strongly damped on a very short time scale.

Starting point of our analysis is the piston equation of motion along the equilibrium line

$$\ddot{X} = -\sqrt{8nRM_g / \pi M^2}\,[(\sqrt{T_1}/X) + \sqrt{T_2}(L-X)]\dot{X}$$

$$+ (M_g / M)[1/X - 1/(L-X)]\dot{X}^2 + A(t) \,, \qquad (5)$$

where A(t) is the fluctuating Langevin's acceleration associated with the random molecular hits suffered by the piston. We note that Eq.(5) differs from Eq.(3) for the presence of A(t) and for the absence of the first term on the R.H.S. of Eq.(3). Actually, this contribution vanishes since the system is on the (equal-pressure) equilibrium line, but the Langevin acceleration A(t) , which was obviously negligible far from equilibrium, can be no longer ignored. Without the term A(t), the structure of Eq.(5) would imply the piston to remain at rest after having reached with zero velocity a position on the equilibrium line obtained by solving the set of Eqs.(1)-(3). Actually, the presence of A(t) is now the main responsible for the stochastic motion of the piston. Correspondingly, our system undergoes a slow random wandering along the states characterized by the common pressure $P_f$ of the two gases (see Fig.2), so that these states should be more properly termed *quasi-equilibrium states*. We wish now to investigate the stochastic piston motion. By exploiting the equilibrium pressure relation $T_1/X=T_2/(L-X)=2T_0/L$ (now valid at all times), Eq.(5) can be rewritten as

$$\ddot{X} = -\sqrt{16nRM_g T_0 / \pi M^2 L} \ (1 / \sqrt{X} + 1 / \sqrt{L - X})\dot{X}$$

$$- (2M_g / M)[(X - L / 2) / X(L - X)]\dot{X}^2 + A(t). \tag{6}$$

Hereafter, we will assume the displacements from $X=L/2$ to be small compared to $L/2$ itself, i.e., $\&X-L/2\&/(L/2)<<1$. Therefore, the coefficients of the terms in $\dot{X}$ and in $\dot{X}^2$ can be approximated by their values at $X=L/2$, so that Eq.(6) can be recast in the form

$$\ddot{X} = -(8 / ML)\sqrt{2nRM_g T_0 / \pi}\dot{X} - (8M_g / ML^2)\dot{X}^2(X - \frac{L}{2}) + A(t) \ . \tag{7}$$

The nonlinear stochastic Eq.(7) contains a friction term opposite to the piston velocity $\dot{X}$ and a term driving the piston toward the particular quasi-equilibrium position $X=L/2$ characterized by maximum entropy.

While solving Eq.(7) is a formidable task, the basic aspects of the piston dynamical evolution can be obtained by approximating the square of the piston velocity with its average thermal value at temperature $(T_A+T_B)/2$, i.e., $\dot{X}^2 = KT_0/M$, where K is Boltzmann's constant. In effect, the restoration term in Eq.(7) does not depend on the sign of the velocity , and the above approximation does not significantly alter its role in determining the most significant consequences of Eq. (7). Proceeding in this way, Eq. (7) is superseded by

$$\ddot{X} = -(8 / ML)\sqrt{2nRM_g T_0 / \pi}\dot{X} - (8KM_g T_0 / M^2 L^2)(X - \frac{L}{2}) + A(t) \ . \tag{8}$$

Equation (8) is in a significant form in that it describes the Brownian motion of a body of mass M (i.e., the piston) in a harmonic field of force driving the piston to the position $X=L/2$, a case which has been exhaustively discussed by Chandrasekhar [4]. More precisely, the author deals with the theory of Brownian motion of a harmonically bound particle of mass m described by the equation $\ddot{x} + \beta\dot{x} + \omega^2 x = A(t)$ (see Eq. (187) of Ref.[4]). By solving the stochastic equation, he is able to prove that $<x^2>_{Av} = KT/m\omega^2$ as $t \to \infty$ (see Eqs. (214) and (215) of Ref.[4]), the characteristic time constant $t_{as}$ over which this averaged square displacement is reached being given by $t_{as} = 1/[\beta - (\beta^2 - 4\omega^2)^{1/2}]$ These results can be transferred to Eq.(8) if we take x=X-L/2, and identify $\beta$ with $(8/ML)[2nRM_gT_0/\pi]^{1/2}$ and $\omega^2$ with $8KM_gT_0/M^2L^2$. After setting $T=T_0$ and $m=M$, we thus obtain

$$< (X - L/2)^2 >_{Av} /(L/2)^2 \rightarrow (M/2M_g) \ , \ \ t \rightarrow \infty \ , \quad (9)$$

for $t > t_{as} \cong \beta/2\omega^2 = ML(2\pi m_g KT_0)^{1/2}$ , where we have taken advantage of the relation $\beta >> \omega$ and $m_g$ is the mass of a single molecule of the gas. In particular, Eq.(9) shows that, since $|X-L/2|$ cannot obviously exceed L/2, we can only deal with the case $M/2M_g<1$ (actually, the hypothesis of small oscillations of the piston around the position X=L/2 , expedient to pass from Eq.(6) to Eq.(7), requires that $M/2M_g<<1$).

We will now give an estimate of the asymptotic entropy decrease if the initial piston position is X(0)=L/2, corresponding to maximum entropy. In order to evaluate the asymptotic value $\Delta S_{as}$ of the entropy change in a rigorous way, one should perform the ensemble average of Eq.(4) with the appropriate asymptotic probability distribution (see Eq.(213) of Ref.[4]). Actually, for the purpose of obtaining an order of magnitude of $\Delta S_{as}$ , it is sufficient to substitute on the R.H.S. of Eq.(4) X and $X^2$ with $<X>_{Av} = L/2$ and $<X^2>_{Av} = L^2(1+M/2M_g)/4$ (see Eq.(9)), which yields

$$\Delta S_{as} \cong nc_p ln(1 - \frac{M}{2M_g}) \cong - nc_p \frac{M}{2M_g} < 0 \ . \quad (10)$$

The quantitative relevance of the above negative entropy variation can be appreciated by comparing it with the standard entropy fluctuations $\Delta S_{th}$ taking place in a gas at thermal equilibrium. To this aim, we recall the well-known result according to which $\Delta S_{th} \equiv (<(\Delta S)^2>_{th})^{1/2}$ is given by $(Knc_p)^{1/2}$ [5], so that we can write, with the help of Eq.(10) and neglecting a factor of the order of unity, $|\Delta S_{as}|/\Delta S_{th} \cong N^{1/2}(M/M_g)$, where N is the total number of gas molecules, which can be a large number.

It is important to explore why the entropy decrease is in our case much larger than that associated with thermal fluctuations. The reason is essentially two-fold:

1. In most cases, thermal equilibrium corresponding to maximum entropy is maintained under the competing influences of equilibrium-disturbing Langevin forces and macroscopic restoration forces driving the system back to equilibrium; for example, if we substitute the adiabatic piston with a thermally conducting one, its random displacements from the central position due to Langevin forces are immediately contrasted by the induced pressure difference between parts A and B. In this respect, our system, in which parts A and B are separated by the adiabatic movable wall, is unique in that the piston displacement from the symmetric position X=L/2 gives rise to a fairly small restoration force (second term on the R.H.S of Eq.(8)), whose weakness is due to the succession of contiguous *quasi-equilibrium* equal-pressure states characterizing the system evolution.

2. H-theorem, which would forbid any substantial entropy decrease, does not apply to our case since one of the main hypotheses underlying this theorem, i.e., the *molecular chaos assumption*, is not fulfilled by our system. As a matter of fact, this hypothesis amounts to assume the validity of the relation $W(\mathbf{r},\mathbf{v};\mathbf{r},\mathbf{v}') = W(\mathbf{r},\mathbf{v})W(\mathbf{r},\mathbf{v}')$ between two-particle and one-particle probability-density functions [1] ; in our case the above equation cannot be satisfied since, while $W(\mathbf{r},\mathbf{v})$ and $W(\mathbf{r},\mathbf{v}')$ are independent from the signs of $v_x$ and $v'_x$ , the correlation induced by the piston random motion favors a common sign of $v_x$ and $v'_x$ in $W(\mathbf{r},\mathbf{v};\mathbf{r},\mathbf{v}')$ for $\mathbf{r}$ close to the piston. In other words, our system does not violate Boltzman's theorem but simply lies outside its realm of applicability, so that no statement can be *a priori* made about the entropy behavior. In fact, while the system entropy decreases if the initial piston position is close to X=L/2 (see Eq.(10)), it is possible to show that it increases if the piston starts from a position X close to zero or to L.

In this paper, we have provided what we believe to be a remarkable example of a physical process violating the second law of thermodynamics which states that "*for any transformation occurring in an isolated system, the entropy of the final state can never be less than that of the initial state* ". Actually, the second law cannot *always* be proved in the frame of Statistical Mechanics. A proof of the second law is furnished by Boltzmann's H-theorem, *provided* the assumptions underlying this theorem are satisfied by the considered isolated system. In other words, H-theorem represents only a sufficient condition for the validity of the Second Law. In our case, since one of the conditions for its applicability (molecular chaos assumption) is not present, there is in principle no reason why the second law could not be violated.

A few comments are necessary in order to justify the model upon which the validity of our result is based. We start by noting that the mass of our piston can be neither too large nor too small compared to that of the gas. In fact, in the first case we would violate the condition $M/2M_g < 1$ (see after Eq.(9)), while in the second the molecular hits would cause the relevant piston displacement to take place on a time scale too short to allow for the thermalization of the gas required in the frame of our model. This condition can be fulfilled if $t_{as} \cong (M/M_g)NL/v_{rms}$ (where $v_{rms}$ is the root mean square velocity of the gas) is larger than the thermalization time $t_{th}$ (which is approximately estimated to be $L^2/\ell v_{rms}$, where $\ell$ is the gas mean free path), that is if $N >> (M_g/M)L/\ell$. On the other hand, the values of N and L are limited by the constrain that $t_{as}$ cannot be so large to become irrelevant. Both requirements can be satisfied for a gas in standard conditions by taking, e.g., $M/M_g = .01$ and L of the order of a few microns. The corresponding $t_{as}$ turns out to be of the order of a tenth of a second and $t_{th}$ of the order of hundreds of nanoseconds,[6] while the value of the relative entropy decrease $|\Delta S_{as}|/\Delta S_{th}$ is of the order of $10^3$.

We wish also to note that our problem is analytically treated in the frame of Langevin approach, which is justified whenever the random force corresponding to A(t) can be considered as *external*, i.e., its properties as a function of time are supposed to be known (see, e.g., [7]). In our case, the characteristic time scale $t_c$ of A(t) is by orders of magnitude smaller than $t_{as}$ and $t_{th}$. Thus, the behavior of A(t) is essentially not influenced by the slow piston dynamics, so that the random forces can be supposed to be known *a priori*, and our system possesses the physical basis for a separation of the acting forces into a mechanical part and a random term with known properties. Accordingly, our system can be modeled by considering the piston as a particle undergoing Brownian motion in a harmonic field, correctly described by the Langevin approach.

We note that in a very recent paper [8] the problem of the time evolution of our system was considered by adopting a different approach based on the introduction of the microscopical Boltzmann equation. The conclusions of Ref.[8] are different from ours, in that the system eventually reaches an asymptotic equilibrium state corresponding to maximum entropy and no violation of the second law is present. We believe that this conclusion strongly depends on molecular chaos assumption, which is expedient for obtaining the results contained in [8]. In fact, this hypothesis inevitably leads, through H-theorem, to entropy maximization. In our model, no *a priori* assumption is made about molecular chaos.

We wish finally to observe that the adiabatic piston problem has been recently mentioned as an illustrative example of thermal evolution whose interpretation may involve significant changes in our way of looking at the second law.[9]

In conclusion, we have investigated a thermal isolated system characterized by a succession of contiguous quasi-equilibrium states, and we have shown how it can undergo a large entropy decrease. At the best of our knowledge, the situation described in this paper constitutes one of the most peculiar examples of thermodynamical evolution of an isolated system outside the realm of Boltzmann's H-theorem.

We wish to thank Noel Corngold for helpful discussions.